\documentclass[aps,prb,onecolumn,reprint,superscriptaddress]{revtex4-1}

\usepackage{graphicx}
\usepackage{amsmath,amsfonts,amssymb}
\usepackage{color}
\usepackage{subfigure}
\usepackage{physics}
\usepackage{dsfont}
\usepackage{hyperref}
\usepackage{url}
\usepackage{leftidx}
\usepackage{textcomp}
\usepackage{gensymb}
\usepackage{esvect}
\usepackage{rotating}

\usepackage{lmodern}

\definecolor{emerald}{rgb}{0.31, 0.78, 0.47}
\definecolor{blue(ncs)}{rgb}{0.0, 0.53, 0.74}
\definecolor{emerald2}{rgb}{0.25, 0.73, 0.57}
\definecolor{benjoscolour}{rgb}{1.00, 0.50, 0.00}
\definecolor{orchid}{rgb}{0.6, 0.2, 0.8}
\usepackage[dvipsnames]{xcolor}

\usepackage{lmodern}

\definecolor{emerald}{rgb}{0.31, 0.78, 0.47}
\definecolor{blue(ncs)}{rgb}{0.0, 0.53, 0.74}
\definecolor{emerald2}{rgb}{0.25, 0.73, 0.57}
\definecolor{benjoscolour}{rgb}{1.00, 0.50, 0.00}
\definecolor{orchid}{rgb}{0.6, 0.2, 0.8}

\hypersetup{
     colorlinks=true,
     linkcolor=blue,
     filecolor=blue,
     citecolor=emerald,      
     urlcolor =blue(ncs),
}

\DeclareMathAlphabet{\pazocal}{OMS}{zplm}{m}{n}

\DeclareMathAlphabet{\pazocal}{OMS}{zplm}{m}{n}

\newcommand{\bo}[1]{\boldsymbol{#1}}
\newcommand{\D}{\mathrm{d}}

\newcommand{\eV}{\mathrm{eV}}
\newcommand{\meV}{\mathrm{meV}}

\newcommand{\e}{\mathrm{e}}

\begin{document}

\title{Comment on ``Axion-matter coupling in multiferroics''}
  
\author{Alexander V. Balatsky}
 \email{balatsky@hotmail.com}
 \affiliation{Nordita, KTH Royal Institute of Technology and Stockholm University, Hannes Alfv\'{e}ns v\"{a}g 12, SE-106 91 Stockholm, Sweden}
  \affiliation{Department of Physics and Institute for Materials Science, University of Connecticut, Storrs, CT 06269, USA}
  
\author{Benjo Fraser}
 \affiliation{Nordita, KTH Royal Institute of Technology and Stockholm University, Hannes Alfv\'{e}ns v\"{a}g 12, SE-106 91 Stockholm, Sweden}

\date{\today}  
  
\begin{abstract}
A previous publication [H.~S.~R{\o}ising et al.,  Phys.~Rev.~Research {\bf{3}}, 033236 (2021)] involving the current authors pointed out a coupling between dark matter axions and ferroic orders in multiferroics. In this comment we argue that using this coupling for dark matter sensing is likely not feasible for the material class we considered, with present-day technologies and level of materials synthesis. The  proposed effect (for QCD axions) is small and is overwhelmed  by thermal noise. This finding means that likely materials for the proposed detection scheme would need to be found with significantly lower magnetic ordering temperatures. 
\end{abstract}
  
\maketitle

In a previous publication~\cite{RoisingEA21} we considered the coupling between dark matter axions and electrons in multiferroics. The coupling was found to yield an energy contribution of the form $g a V (\mu_0 / \varepsilon_0)^{1/2} \bo{P} \cdot \bo{M}$, where $\bo{P}$ ($\bo{M}$) is the ferroelectric (ferromagnetic) polarization vector, $V = L_{\mathrm{domain}}^3$ is the volume of the homogeneous ferroic domains, $a$ is the axion field, and $g \sim 10^{-10} g_{ae}$ where $g_{ae}$ is the bare axion-electron coupling. A linear response estimate suggested that the coupling could lead to a time-dependent magnetic response on the order of $\delta M \sim \pazocal{O}(1\mathrm{aT})$ under ideal conditions with parameters motivated by hexagonal  Lu$_{1-x}$Sc$_x$FeO$_3$ ($h$-LSFO), a candidate $\bo{P} \parallel \bo{M}$ multiferroic. We suggested that multiferroics therefore might be a platform for sensing dark matter axions using hypersensitive magnetometers and macroscopic sensor volumes. In this comment, we provide order-of-magnitude noise estimates suggesting that mK temperatures and $V \sim 1\mathrm{m}^3$ may be required to achieve a signal-to-noise ratio greater than one. These tight temperature and volume constraints would make it challenging to sense dark matter axions in multiferroics with present-day technologies and existing material candidates.

In Ref.~\onlinecite{RoisingEA21} we used a Ginzburg-Landau model for the longitudinal magnetization perturbation $\delta M(t)$ induced by the axion. In this note, we extend the model to include the effects of noise by adding a stochastic term to the equations of motion. This produces the Langevin equation
\begin{equation}
\begin{aligned}
\frac{\D^2\delta M(t)}{\D t^2} + &\gamma \frac{\D\delta M(t)}{\D t} + m_M^2\delta M(t) \\
&= P_0 \theta(t) + \xi(t),
\end{aligned}
\label{eq:Langevin}
\end{equation}
where $\xi(t)$ is the noise term, $\gamma$ the damping factor (the width of the magnetic resonance), $P_0$ the static ferroelectric polarization, and $\theta(t) = \theta_0 a(t)$ is the axion driving term. We allow for the possibility of a bandwidth $\Delta\omega_a$ for the axion signal. Solving \eqref{eq:Langevin}, the power spectral density $S_M(\omega) \equiv \int\dd t\, \e^{i \omega t}\langle \delta M(t) \delta M(0) \rangle$ is equal to
\begin{align}
S_M (\omega)\, &=\, |\chi_M(\omega)|^2\left[S_\theta(\omega) + S_\xi(\omega)\right]\\
\chi_M(\omega)\, &=\, \frac{1}{-(\omega^2-m_M^2)- i\gamma \omega}
\label{eq:spectraldensity}
\end{align}
where $\chi_M(\omega)$ is the response function of equation \eqref{eq:Langevin}, and $S_\theta(\omega)$, $S_\xi(\omega)$ are the spectral densities of the two driving terms on its right hand side. 

We assume white noise with correlation function $\langle\xi(t)\xi(t')\rangle=\lambda\, \delta(t-t')$. The kinetic energy of the magnetization in the Ginzburg-Landau model of~\onlinecite{RoisingEA21} is $F(M) = \int \dd^3 x\frac{1}{2}\alpha_M(\partial_t M)^2$, where we guess $\alpha_M=(1 \meV)^{-2}$ based on typical spin-exchange couplings; the value of this constant has not been directly measured for $h$-LSFO. Then the fluctuation-dissipation theorem gives $\lambda \sim T\gamma/(V\alpha_M)$ in the classically limited case (we are in this limit since the LSFO Curie temperature is $T\sim 100K\gg\hbar\omega_a/k_B$). 

There are three relevant bandwidths in the problem: 
\begin{itemize}
    \item $\gamma \sim 10^{-6}\eV$ the width of the magnetic resonance: our numerical estimate here are based on inelastic neutron scattering measurements~\cite{PhysRevB.98.134412}
    \item $\Delta\omega_a \sim \frac 12\omega_a v_a^2 \sim 10^{-12}\eV$ the width of the axion signal, expected to be determined by Doppler broadening of the Galactic axion background 
    \item $\Delta\omega_{\mathrm{meas}}\sim t_{\mathrm{meas}}^{-1}\sim 10^{-18}\eV$ the measurement bandwidth, which is determined by the measurement time $t_{\mathrm{meas}}$: this estimate assumes $t_{\mathrm{meas}}\sim 1 \text{hr}$.
\end{itemize}
We therefore see that $\gamma\gg \Delta\omega_a\gg \Delta\omega_{\mathrm{meas}}$. In this regime the signal-to-noise ratio is~\cite{BudkerEA14, Sikivie20}
\begin{align}
    \frac{S}{N}\, =\, \frac{S_\theta(\omega_a)}{S_\xi(\omega_a)} \sqrt{\frac{\Delta\omega_a}{\Delta\omega_{\mathrm{meas}}}}
\end{align}
where the square root factor does not come from \eqref{eq:spectraldensity}, but from that fact that we can take $N\sim \Delta\omega_a/\Delta\omega_{\mathrm{meas}}$ samples by scanning across the axion resonance - see the Appendix of \onlinecite{BudkerEA14}. This is the regime where the measurement time is much greater than the axion coherence time, so that the amplitude signal-to-noise ratio $\sqrt{\frac{S}{N}}\propto t_{\mathrm{meas}}^{1/4}$. 

For our case we find the order-of-magnitude estimate
\begin{equation}
\begin{aligned}
    \frac{S}{N}\, &\sim\, \frac{V}{\gamma T} \frac{|P_0|^2}{\Delta\omega_a\alpha_M}(\chi\theta)^2\sqrt{\frac{\Delta\omega_a}{\Delta\omega_{\mathrm{meas}}}}\\
    &= 10^{-8} \left( \frac{V}{1 \mathrm{m}^3} \right)\left( \frac{10^{-5}\eV}{\gamma} \right)\left( \frac{100\mathrm{K}}{T} \right)\sqrt{\frac{\Delta\omega_a}{\Delta\omega_{\mathrm{meas}}}}
\end{aligned}
\label{eq:SNBenjo}
\end{equation}
\begin{figure}
	\centering
	 \includegraphics[width=0.5\textwidth]{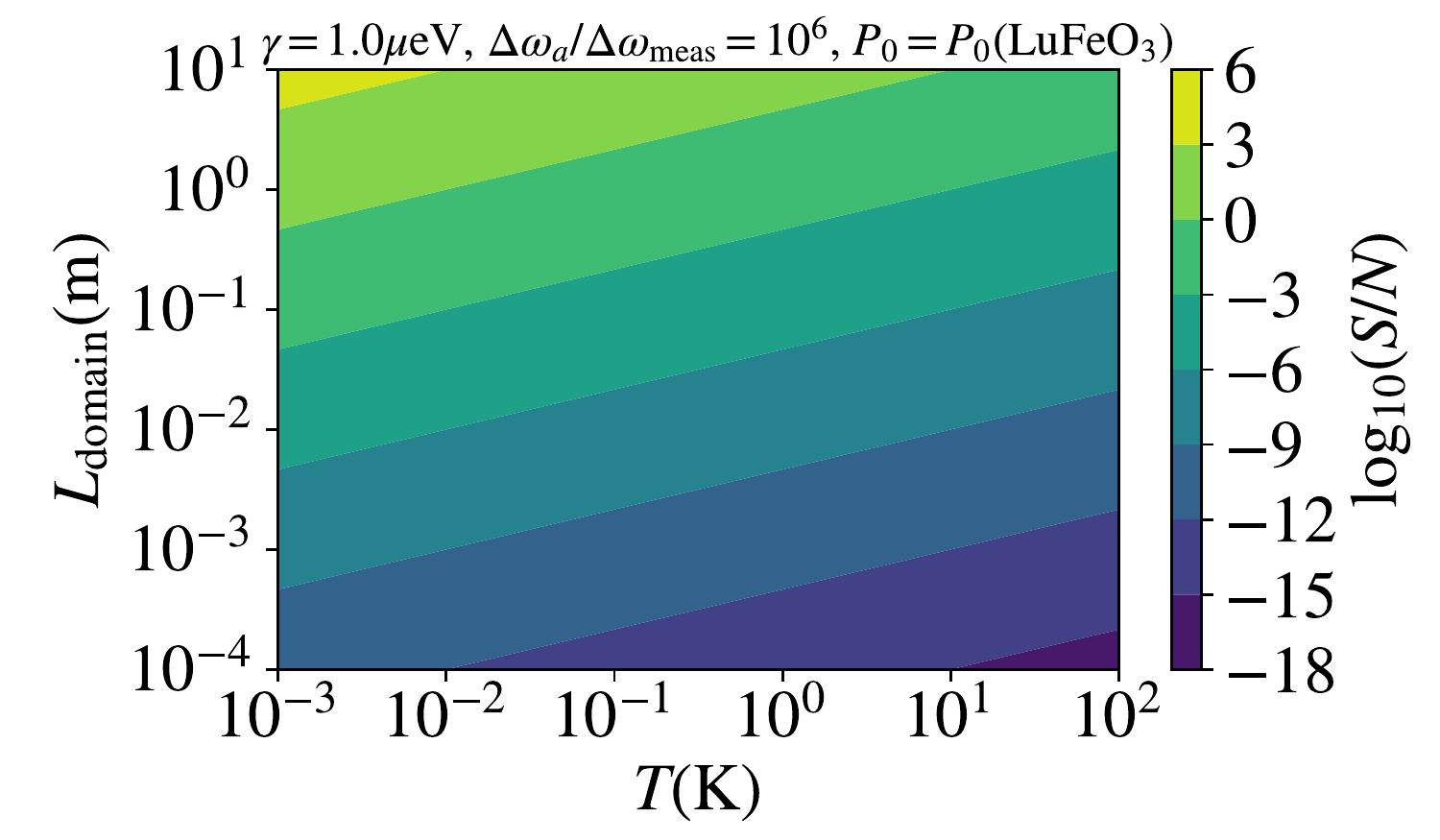}
	\caption{Contour plot for the estimated signal-to-noise ratios from Eq.~\eqref{eq:SNBenjo}. These estimates assume the noise to be classically limited, and as reference parameters we use those reported in Ref.~\onlinecite{RoisingEA21}, which are motivated by $h$-LSFO.}
	\label{fig:Noise}
\end{figure}
In Fig.~\ref{fig:Noise} we plot the case of classically limited noise as a function of the linear size of the homogeneous domains $L_{\mathrm{domain}} = V^{1/3}$. Untrained samples of $h$-LSFO display ferroelectric (ferromagnetic) domains of typical size $10\mu$m ($100\mu$m)~\cite{DuEA18}, the former of which can be controlled by the quench rate through the transition~\cite{Griffin12}. Training techniques can in ideal cases push the homogeneous coupling domains to the order of $1$mm~\footnote{Private communication with Sang-Wook Cheong}, which from the estimate of Fig.~\ref{fig:Noise} is four orders of magnitude short from achieving a signal-to-noise ratio greater than one, even for mK temperatures. 

The absolute value of the magnetic signal, which our estimates suggest to be on the order of $\pazocal{O}(1~\mathrm{aT})$ for $h$-LSFO motivated parameters, is at the lower end of what can feasibly be measured with present-day SQUID technologies having sensitivities on the order of $10^{-16}~\mathrm{T}/\sqrt{\mathrm{Hz}}$. The Gravity Probe B experiment~\cite{EverittEA11} demonstrated sensitivities to magnetic field deviations of about $5~\mathrm{aT}$, with sampling times of a few days. 

These small numbers for our proposed multiferroic indicate (as it is currently synthesized) that it is unsuitable for detection of the QCD axion through our mechanism. However, we do not rule out the mechanism's future viability, for example by optimizing the material properties to improve the signal-to-noise ratio. We note that the driving term of the axion-induced perturbation is proportional to the ferroelectric polarization, so one could search for multiferroics where the polarization is large, including hybrid structures and field-induced sensing devices. Lone pair ferroelectrics, such as the archetypical BiFeO$_3$, have the potential to reach polarizations at least an order of magnitude greater than $h$-LSFO~\cite{FiebigEA16}. 

Finally, we mention that should an alternative mechanism be found in which the axion couples to the matter fields $\bo{P}$ and $\bo{M}$ directly from the axion-photon coupling, $\frac{a}{f_a} F_{\mu \nu} \tilde{F}^{\mu \nu}$, this could improve significantly on some of the above issues, since such a coupling would not suffer from the $m_a / m_e$ suppression that enters in the the axion-fermion coupling $\frac{\partial_{\mu} a}{f_a} \bar{\Psi}_f \gamma^{\mu} \gamma^5 \Psi_f$ we considered here. For an $\mu\eV$ mass axion this ratio is of the order $10^{-10}$!

{\em Acknowledgements}: This work developed from our  discussions with Henrik S. R{\o}ising, to whom we are grateful for analysis, comments and critique. We also acknowledge discussions with S.W. Cheong, J. Conrad, S. Griffin, N. Spaldin and F. Wilczek.   This work was funded by VR  Axion research environment grant `Detecting Axion Dark Matter In The Sky And In The Lab (AxionDM)' funded by the Swedish Research Council (VR) under Dnr 2019-02337, European Research Council ERC HERO-810451 grant and University of Connecticut.

\bibliographystyle{apsrev4-1}
\bibliography{MultiferroicsDM}

\begin{thebibliography}{9}%
\makeatletter
\providecommand \@ifxundefined [1]{%
 \@ifx{#1\undefined}
}%
\providecommand \@ifnum [1]{%
 \ifnum #1\expandafter \@firstoftwo
 \else \expandafter \@secondoftwo
 \fi
}%
\providecommand \@ifx [1]{%
 \ifx #1\expandafter \@firstoftwo
 \else \expandafter \@secondoftwo
 \fi
}%
\providecommand \natexlab [1]{#1}%
\providecommand \enquote  [1]{``#1''}%
\providecommand \bibnamefont  [1]{#1}%
\providecommand \bibfnamefont [1]{#1}%
\providecommand \citenamefont [1]{#1}%
\providecommand \href@noop [0]{\@secondoftwo}%
\providecommand \href [0]{\begingroup \@sanitize@url \@href}%
\providecommand \@href[1]{\@@startlink{#1}\@@href}%
\providecommand \@@href[1]{\endgroup#1\@@endlink}%
\providecommand \@sanitize@url [0]{\catcode `\\12\catcode `\$12\catcode
  `\&12\catcode `\#12\catcode `\^12\catcode `\_12\catcode `\%12\relax}%
\providecommand \@@startlink[1]{}%
\providecommand \@@endlink[0]{}%
\providecommand \url  [0]{\begingroup\@sanitize@url \@url }%
\providecommand \@url [1]{\endgroup\@href {#1}{\urlprefix }}%
\providecommand \urlprefix  [0]{URL }%
\providecommand \Eprint [0]{\href }%
\providecommand \doibase [0]{http://dx.doi.org/}%
\providecommand \selectlanguage [0]{\@gobble}%
\providecommand \bibinfo  [0]{\@secondoftwo}%
\providecommand \bibfield  [0]{\@secondoftwo}%
\providecommand \translation [1]{[#1]}%
\providecommand \BibitemOpen [0]{}%
\providecommand \bibitemStop [0]{}%
\providecommand \bibitemNoStop [0]{.\EOS\space}%
\providecommand \EOS [0]{\spacefactor3000\relax}%
\providecommand \BibitemShut  [1]{\csname bibitem#1\endcsname}%
\let\auto@bib@innerbib\@empty
\bibitem [{\citenamefont {R\o{}ising}\ \emph {et~al.}(2021)\citenamefont
  {R\o{}ising}, \citenamefont {Fraser}, \citenamefont {Griffin}, \citenamefont
  {Bandyopadhyay}, \citenamefont {Mahabir}, \citenamefont {Cheong},\ and\
  \citenamefont {Balatsky}}]{RoisingEA21}%
  \BibitemOpen
  \bibfield  {author} {\bibinfo {author} {\bibfnamefont {H.~S.}\ \bibnamefont
  {R\o{}ising}}, \bibinfo {author} {\bibfnamefont {B.}~\bibnamefont {Fraser}},
  \bibinfo {author} {\bibfnamefont {S.~M.}\ \bibnamefont {Griffin}}, \bibinfo
  {author} {\bibfnamefont {S.}~\bibnamefont {Bandyopadhyay}}, \bibinfo {author}
  {\bibfnamefont {A.}~\bibnamefont {Mahabir}}, \bibinfo {author} {\bibfnamefont
  {S.-W.}\ \bibnamefont {Cheong}}, \ and\ \bibinfo {author} {\bibfnamefont
  {A.~V.}\ \bibnamefont {Balatsky}},\ }\href {\doibase
  10.1103/PhysRevResearch.3.033236} {\bibfield  {journal} {\bibinfo  {journal}
  {Phys. Rev. Research}\ }\textbf {\bibinfo {volume} {3}},\ \bibinfo {pages}
  {033236} (\bibinfo {year} {2021})}\BibitemShut {NoStop}%
\bibitem [{\citenamefont {Leiner}\ \emph {et~al.}(2018)\citenamefont {Leiner},
  \citenamefont {Kim}, \citenamefont {Park}, \citenamefont {Oh}, \citenamefont
  {Perring}, \citenamefont {Walker}, \citenamefont {Xu}, \citenamefont {Wang},
  \citenamefont {Cheong},\ and\ \citenamefont {Park}}]{PhysRevB.98.134412}%
  \BibitemOpen
  \bibfield  {author} {\bibinfo {author} {\bibfnamefont {J.~C.}\ \bibnamefont
  {Leiner}}, \bibinfo {author} {\bibfnamefont {T.}~\bibnamefont {Kim}},
  \bibinfo {author} {\bibfnamefont {K.}~\bibnamefont {Park}}, \bibinfo {author}
  {\bibfnamefont {J.}~\bibnamefont {Oh}}, \bibinfo {author} {\bibfnamefont
  {T.~G.}\ \bibnamefont {Perring}}, \bibinfo {author} {\bibfnamefont {H.~C.}\
  \bibnamefont {Walker}}, \bibinfo {author} {\bibfnamefont {X.}~\bibnamefont
  {Xu}}, \bibinfo {author} {\bibfnamefont {Y.}~\bibnamefont {Wang}}, \bibinfo
  {author} {\bibfnamefont {S.-W.}\ \bibnamefont {Cheong}}, \ and\ \bibinfo
  {author} {\bibfnamefont {J.-G.}\ \bibnamefont {Park}},\ }\href {\doibase
  10.1103/PhysRevB.98.134412} {\bibfield  {journal} {\bibinfo  {journal} {Phys.
  Rev. B}\ }\textbf {\bibinfo {volume} {98}},\ \bibinfo {pages} {134412}
  (\bibinfo {year} {2018})}\BibitemShut {NoStop}%
\bibitem [{\citenamefont {Budker}\ \emph {et~al.}(2014)\citenamefont {Budker},
  \citenamefont {Graham}, \citenamefont {Ledbetter}, \citenamefont
  {Rajendran},\ and\ \citenamefont {Sushkov}}]{BudkerEA14}%
  \BibitemOpen
  \bibfield  {author} {\bibinfo {author} {\bibfnamefont {D.}~\bibnamefont
  {Budker}}, \bibinfo {author} {\bibfnamefont {P.~W.}\ \bibnamefont {Graham}},
  \bibinfo {author} {\bibfnamefont {M.}~\bibnamefont {Ledbetter}}, \bibinfo
  {author} {\bibfnamefont {S.}~\bibnamefont {Rajendran}}, \ and\ \bibinfo
  {author} {\bibfnamefont {A.~O.}\ \bibnamefont {Sushkov}},\ }\href {\doibase
  10.1103/PhysRevX.4.021030} {\bibfield  {journal} {\bibinfo  {journal} {Phys.
  Rev. X}\ }\textbf {\bibinfo {volume} {4}},\ \bibinfo {pages} {021030}
  (\bibinfo {year} {2014})}\BibitemShut {NoStop}%
\bibitem [{\citenamefont {Sikivie}(2021)}]{Sikivie20}%
  \BibitemOpen
  \bibfield  {author} {\bibinfo {author} {\bibfnamefont {P.}~\bibnamefont
  {Sikivie}},\ }\href {\doibase 10.1103/RevModPhys.93.015004} {\bibfield
  {journal} {\bibinfo  {journal} {Rev. Mod. Phys.}\ }\textbf {\bibinfo {volume}
  {93}},\ \bibinfo {pages} {015004} (\bibinfo {year} {2021})}\BibitemShut
  {NoStop}%
\bibitem [{\citenamefont {Du}\ \emph {et~al.}(2018)\citenamefont {Du},
  \citenamefont {Gao}, \citenamefont {Wang}, \citenamefont {Xu}, \citenamefont
  {Kim}, \citenamefont {Hu}, \citenamefont {Huang},\ and\ \citenamefont
  {Cheong}}]{DuEA18}%
  \BibitemOpen
  \bibfield  {author} {\bibinfo {author} {\bibfnamefont {K.}~\bibnamefont
  {Du}}, \bibinfo {author} {\bibfnamefont {B.}~\bibnamefont {Gao}}, \bibinfo
  {author} {\bibfnamefont {Y.}~\bibnamefont {Wang}}, \bibinfo {author}
  {\bibfnamefont {X.}~\bibnamefont {Xu}}, \bibinfo {author} {\bibfnamefont
  {J.}~\bibnamefont {Kim}}, \bibinfo {author} {\bibfnamefont {R.}~\bibnamefont
  {Hu}}, \bibinfo {author} {\bibfnamefont {F.-T.}\ \bibnamefont {Huang}}, \
  and\ \bibinfo {author} {\bibfnamefont {S.-W.}\ \bibnamefont {Cheong}},\
  }\href {\doibase 10.1038/s41535-018-0106-3} {\bibfield  {journal} {\bibinfo
  {journal} {npj Quantum Mater.}\ }\textbf {\bibinfo {volume} {3}},\ \bibinfo
  {pages} {33} (\bibinfo {year} {2018})}\BibitemShut {NoStop}%
\bibitem [{\citenamefont {Griffin}\ \emph {et~al.}(2012)\citenamefont
  {Griffin}, \citenamefont {Lilienblum}, \citenamefont {Delaney}, \citenamefont
  {Kumagai}, \citenamefont {Fiebig},\ and\ \citenamefont
  {Spaldin}}]{Griffin12}%
  \BibitemOpen
  \bibfield  {author} {\bibinfo {author} {\bibfnamefont {S.~M.}\ \bibnamefont
  {Griffin}}, \bibinfo {author} {\bibfnamefont {M.}~\bibnamefont {Lilienblum}},
  \bibinfo {author} {\bibfnamefont {K.~T.}\ \bibnamefont {Delaney}}, \bibinfo
  {author} {\bibfnamefont {Y.}~\bibnamefont {Kumagai}}, \bibinfo {author}
  {\bibfnamefont {M.}~\bibnamefont {Fiebig}}, \ and\ \bibinfo {author}
  {\bibfnamefont {N.~A.}\ \bibnamefont {Spaldin}},\ }\href {\doibase
  10.1103/PhysRevX.2.041022} {\bibfield  {journal} {\bibinfo  {journal} {Phys.
  Rev. X}\ }\textbf {\bibinfo {volume} {2}},\ \bibinfo {pages} {041022}
  (\bibinfo {year} {2012})}\BibitemShut {NoStop}%
\bibitem [{Note1()}]{Note1}%
  \BibitemOpen
  \bibinfo {note} {Private communication with Sang-Wook Cheong}\BibitemShut
  {NoStop}%
\bibitem [{\citenamefont {Everitt}\ \emph {et~al.}(2011)\citenamefont
  {Everitt}, \citenamefont {DeBra}, \citenamefont {Parkinson}, \citenamefont
  {Turneaure}, \citenamefont {Conklin}, \citenamefont {Heifetz}, \citenamefont
  {Keiser}, \citenamefont {Silbergleit}, \citenamefont {Holmes}, \citenamefont
  {Kolodziejczak}, \citenamefont {Al-Meshari}, \citenamefont {Mester},
  \citenamefont {Muhlfelder}, \citenamefont {Solomonik}, \citenamefont {Stahl},
  \citenamefont {Worden}, \citenamefont {Bencze}, \citenamefont {Buchman},
  \citenamefont {Clarke}, \citenamefont {Al-Jadaan}, \citenamefont
  {Al-Jibreen}, \citenamefont {Li}, \citenamefont {Lipa}, \citenamefont
  {Lockhart}, \citenamefont {Al-Suwaidan}, \citenamefont {Taber},\ and\
  \citenamefont {Wang}}]{EverittEA11}%
  \BibitemOpen
  \bibfield  {author} {\bibinfo {author} {\bibfnamefont {C.~W.~F.}\
  \bibnamefont {Everitt}}, \bibinfo {author} {\bibfnamefont {D.~B.}\
  \bibnamefont {DeBra}}, \bibinfo {author} {\bibfnamefont {B.~W.}\ \bibnamefont
  {Parkinson}}, \bibinfo {author} {\bibfnamefont {J.~P.}\ \bibnamefont
  {Turneaure}}, \bibinfo {author} {\bibfnamefont {J.~W.}\ \bibnamefont
  {Conklin}}, \bibinfo {author} {\bibfnamefont {M.~I.}\ \bibnamefont
  {Heifetz}}, \bibinfo {author} {\bibfnamefont {G.~M.}\ \bibnamefont {Keiser}},
  \bibinfo {author} {\bibfnamefont {A.~S.}\ \bibnamefont {Silbergleit}},
  \bibinfo {author} {\bibfnamefont {T.}~\bibnamefont {Holmes}}, \bibinfo
  {author} {\bibfnamefont {J.}~\bibnamefont {Kolodziejczak}}, \bibinfo {author}
  {\bibfnamefont {M.}~\bibnamefont {Al-Meshari}}, \bibinfo {author}
  {\bibfnamefont {J.~C.}\ \bibnamefont {Mester}}, \bibinfo {author}
  {\bibfnamefont {B.}~\bibnamefont {Muhlfelder}}, \bibinfo {author}
  {\bibfnamefont {V.~G.}\ \bibnamefont {Solomonik}}, \bibinfo {author}
  {\bibfnamefont {K.}~\bibnamefont {Stahl}}, \bibinfo {author} {\bibfnamefont
  {P.~W.}\ \bibnamefont {Worden}}, \bibinfo {author} {\bibfnamefont
  {W.}~\bibnamefont {Bencze}}, \bibinfo {author} {\bibfnamefont
  {S.}~\bibnamefont {Buchman}}, \bibinfo {author} {\bibfnamefont
  {B.}~\bibnamefont {Clarke}}, \bibinfo {author} {\bibfnamefont
  {A.}~\bibnamefont {Al-Jadaan}}, \bibinfo {author} {\bibfnamefont
  {H.}~\bibnamefont {Al-Jibreen}}, \bibinfo {author} {\bibfnamefont
  {J.}~\bibnamefont {Li}}, \bibinfo {author} {\bibfnamefont {J.~A.}\
  \bibnamefont {Lipa}}, \bibinfo {author} {\bibfnamefont {J.~M.}\ \bibnamefont
  {Lockhart}}, \bibinfo {author} {\bibfnamefont {B.}~\bibnamefont
  {Al-Suwaidan}}, \bibinfo {author} {\bibfnamefont {M.}~\bibnamefont {Taber}},
  \ and\ \bibinfo {author} {\bibfnamefont {S.}~\bibnamefont {Wang}},\ }\href
  {\doibase 10.1103/PhysRevLett.106.221101} {\bibfield  {journal} {\bibinfo
  {journal} {Phys. Rev. Lett.}\ }\textbf {\bibinfo {volume} {106}},\ \bibinfo
  {pages} {221101} (\bibinfo {year} {2011})}\BibitemShut {NoStop}%
\bibitem [{\citenamefont {Fiebig}\ \emph {et~al.}(2016)\citenamefont {Fiebig},
  \citenamefont {Lottermoser}, \citenamefont {Meier},\ and\ \citenamefont
  {Trassin}}]{FiebigEA16}%
  \BibitemOpen
  \bibfield  {author} {\bibinfo {author} {\bibfnamefont {M.}~\bibnamefont
  {Fiebig}}, \bibinfo {author} {\bibfnamefont {T.}~\bibnamefont {Lottermoser}},
  \bibinfo {author} {\bibfnamefont {D.}~\bibnamefont {Meier}}, \ and\ \bibinfo
  {author} {\bibfnamefont {M.}~\bibnamefont {Trassin}},\ }\href {\doibase
  10.1038/natrevmats.2016.46} {\bibfield  {journal} {\bibinfo  {journal} {Nat.
  Rev. Mater.}\ }\textbf {\bibinfo {volume} {1}},\ \bibinfo {pages} {16046}
  (\bibinfo {year} {2016})}\BibitemShut {NoStop}%
\end{thebibliography}%

\end{document}